\documentclass[3p,times,twocolumn]{elsarticle}

\usepackage{ecrc}

\volume{00}

\firstpage{1}

\journalname{Nuclear Physics B Proceedings Supplement}

\runauth{}

\jid{nuphbp}

\jnltitlelogo{Nuclear Physics B Proceedings Supplement}

\usepackage{amssymb}
\usepackage{amsmath}

\usepackage[figuresright]{rotating}

\begin{document}

\begin{frontmatter}


\ead{daristizabal@ulg.ac.be}

\dochead{}

\title{Scalar triplet leptogenesis without right-handed neutrino
    decoupling}


\author{D. Aristizabal Sierra}

\address{IFPA, Dep. AGO, Universite de Liege, Bat B5, Sart Tilman
  B-4000 Liege 1, Belgium}

\begin{abstract}
  We discuss leptogenesis in the context of type-II seesaw in the case
  in which in addition to the scalar electroweak triplet decays the
  lepton asymmetry is also induced by right-handed neutrino decays
  (mild hierarchical scenarios). We show that within this setup,
  depending on the relative sizes of the relevant parameters, one can
  identify three classes of generic models, each one with its own
  consequences for leptogenesis.
\end{abstract}


\end{frontmatter}


\section{Introduction}
\label{sec:intro}
Going beyond standard leptogenesis means going beyond type-I
seesaw. Certainly the most simple frameworks for such a task are
provided by type-II seesaw, type-III seesaw or interplays among them
(a full analysis can be found in \cite{Hambye:2012fh}).  In particular
the ``hybrid'' type-I plus type-II scenario has been discussed in the
limit of right-handed (RH) neutrino
decoupling\cite{Hambye:2005tk}\footnote{Leptogenesis in type-III
  scenarios was discussed for the first time in \cite{Hambye:2003rt}
  and subsequently discussed in \cite{AristizabalSierra:2010mv}.}, so
we here aim to extend upon this consideration by discussing scenarios
were a net non-zero baryon asymmetry is builded up from the dynamics
of both, the electroweak triplet and the lightest RH neutrino,
scenarios we dub as {\it mild hierarchical scenarios}. 

This paper is mostly based on references
\cite{AristizabalSierra:2011ab,AristizabalSierra:2012js}.
\section{Type II seesaw and leptogenesis}
\label{sec:setup-Beqs}
Extending the standard model with a scalar electroweak triplet induces
a new set of interactions determined by
\begin{equation}
  \label{eq:Lag-II}
    {\cal L}^{II}=-Y_{ij}\ell_{L_i}^TCi\tau_2\boldsymbol{\Delta}\ell_{L_j}
    -M^2_\Delta \mbox{Tr}\boldsymbol{\Delta}^\dagger\boldsymbol{\Delta}
  +\mu H^Ti\tau_2\boldsymbol{\Delta} H + \mbox{H.c.}\,,
\end{equation}
with $\boldsymbol{\Delta}$ given by
\begin{equation}
  \label{eq:triplet}
  \boldsymbol{\Delta}=
  \begin{pmatrix}
    \Delta^{++} & \Delta^{+}/\sqrt{2}\\
    \Delta^{+}/\sqrt{2} & \Delta^{0}
  \end{pmatrix}\,.
\end{equation}
These interactions solely can account for neutrino masses and mixings,
with the effective light neutrino mass matrix given by
\begin{equation}
  \label{eq:nmm-II}
  \boldsymbol{m_\nu^{II}}=2\,v_\Delta\,\boldsymbol{Y}\,,
\end{equation}
where $\langle\boldsymbol{\Delta}\rangle=v_\Delta$. Successful
leptogenesis, however, requires going beyond the interactions in
(\ref{eq:Lag-II}). Here we will consider RH neutrinos as the new
degrees of freedom allowing the generation of a baryon asymmetry
consistent with data {\it i.e.} scenarios with interplay between
type-I and type-II seesaw. The interactions induced by the RH
neutrinos,
\begin{equation}
  \label{eq:Lag-I}
    {\cal L}^{I}= -\lambda_{ij}\overline N_{R_i} \ell_{L_j}
  \tilde H^\dagger
  -\frac{1}{2}\overline N_{R_i}C M_{R_i}
  \overline N_{R_i}^T + \mbox{H.c.}\,,
\end{equation}
lead to the standard type-I CP asymmetry parameter in $N_k$ decays,
$\epsilon_{N_k}$, while the combination of interactions in
\eqref{eq:Lag-II} and \eqref{eq:Lag-I} yield a new contribution,
$\epsilon_{N_k}^\Delta$, to the total CP asymmetry,
$\epsilon_{N_k}^\text{tot}$ \cite{Hambye:2003ka}.  While $N_k$ decays
receive contributions from wave-function as well as vertex one-loop
corrections, in the setup considered here triplet decays are subject
only to vertex corrections. Thus, the CP asymmetry parameter in
triplet decays, $\epsilon_\Delta$, arises solely from the interference
between the tree-level and vertex correction.
\subsection{Kinetic equations for triplet and RH neutrino dynamics}
\label{sec:BEqs}
Depending on the mass spectrum of the mixed type-I plus type-II
scenario, the lepton asymmetry can be produced by RH neutrino decays,
triplet decays or both. When writing the Boltzmann equations
accounting for the dynamics of the state producing the lepton
asymmetry those states which are decoupled can be ignored and---in
general---only the lightest state interactions have to be included. In
{\it mild hierarchical scenarios}, where $M_\Delta\sim M_{N_1}$, both the
triplet and lightest RH neutrino reactions play a crucial role.

The full network of kinetic equations for the case under consideration
consist of five coupled differential equations for the following
densities: $Y_{N_1}$, $Y_{\Sigma}$,
$Y_{\Delta_L}$, $Y_{\Delta_H}$, $Y_{\Delta_\Delta}$. The resulting
equations satisfy the constraint
\begin{equation}
  \label{eq:hypercharge-neutrality}
  2\,Y_{\Delta_\Delta} + Y_{\Delta H} - Y_{\Delta_L}=0
\end{equation}
as a consequence of hypercarge conservation\footnote{The same relation
  is found to hold in the case when the RH neutrino is decoupled
  \cite{Hambye:2005tk}. This is expected since the RH neutrino is a
  vanishing hypercharge state.}. With this constraint at hand and in
the one-flavor approximation the final network of Boltzmann equations
can be written according to \cite{AristizabalSierra:2011ab}
\begin{align}
  \label{eq:relevant-BEQs}
  \dot Y_{N_1}=&-(y_{N_1}-1)\,\gamma_{D_{N_1}}\,,
  \nonumber\\
  \dot Y_\Sigma=&-(y_\Sigma - 1)\,\gamma_{D_\Delta} 
  - 2(y_\Sigma^2 - 1)\,\gamma_A\,,\nonumber\\
  \dot Y_{\Delta_L}=&\left[(y_{N_1}-1)\,\epsilon_{N_1}^\text{tot} -
    \left(y_{\Delta_L} - y_{\Delta_\Delta}^H\right)\right]\,\gamma_{D_{N_1}}
  \nonumber\\
  &+\left[(y_\Sigma - 1)\,\epsilon_\Delta -
    2K_\ell\, (y_{\Delta_L} + y_{\Delta_\Delta})\right]\,\gamma_{D_\Delta}\,,
  \nonumber\\
  \dot Y_{\Delta_\Delta}=&-\left[y_{\Delta_\Delta} + (K_\ell - K_H)\,y_{\Delta_L}
    + 2K_H \,y_{\Delta_\Delta}^H\right]\,,
\end{align}
where the following conventions are used: $\dot Y\equiv sHz\,dY/dz$,
$y_X\equiv Y_X/Y_X^{\text{Eq}}$ (the exception being
$y_{\Delta_\Delta}^H\equiv Y_{\Delta_\Delta}/Y_H^\text{Eq}$ and
$y_{\Delta_L}=Y_{\Delta_L}/Y^\text{Eq}_\ell$), $\Sigma\equiv\Delta +
\Delta^\dagger$,
$\epsilon_{N_1}^\text{tot}\equiv\epsilon_{N_1}+\epsilon_{N_1}^\Delta$
and $\gamma_{D_{N_1}}$, $\gamma_{D_\Delta}$ and $\gamma_A$ are the
reaction densities for: RH neutrino and triplet decays and triplet
annihilations.  The factors $K_{\ell,H}$ resemble the flavor
projectors defined in standard flavored leptogenesis
\cite{Nardi:2006fx,Abada:2006fw} as they project triplet decays into
the Higgs or the lepton doublet directions. They are given by
\begin{equation}
  \label{eq:projectors}
  K_\ell=\frac{\tilde m_\Delta^\ell}{\tilde m_\Delta^\ell + \frac{\tilde m_\Delta^2}
    {4\,\tilde m_\Delta^\ell}}\,
  \quad\mbox{and}\quad
  K_H=\frac{\tilde m_\Delta^2}{4\,\tilde m_\Delta^\ell\left(
    \tilde m_\Delta^\ell + \frac{\tilde m_\Delta^2}
    {4\,\tilde m_\Delta^\ell}\right)}\,,
\end{equation}
where the parameters $\tilde m_\Delta^\ell$ and $\tilde m^2_\Delta$ are given by
\begin{equation}
  \label{eq:definition-mtildes}
  \tilde m_\Delta^\ell=\frac{v^2\,|\boldsymbol{Y}|^2}{M_\Delta}
  \quad\mbox{and}\quad
  \tilde m^2_\Delta=\mbox{Tr}[\boldsymbol{m_\nu^{II}}\boldsymbol{m_\nu^{II}}^\dagger]\,,
\end{equation}
with $v=\langle H\rangle\simeq 174$ GeV and
$|\boldsymbol{Y}|^2=\mbox{Tr}[\boldsymbol{Y}\,\boldsymbol{Y}^\dagger]$.

Since we are dealing with a {\it mild hierarchical scenario} we asumme
$r=M_\Delta/M_{N_1}\subset [10^{-1},1]$ and so while $z=M_\Delta/T$,
$z_N=r\,z$. Accordingly, the problem of studying the generation of a
lepton asymmetry through eqs. \eqref{eq:relevant-BEQs}, once the CP
asymmetries $\epsilon_{N_1}^\text{tot}$ and $\epsilon_\Delta$ are
fixed, reduces to a five parameter problem, namely $\tilde
m_{N_1}=v^2(\pmb{\lambda}\pmb{\lambda}^\dagger)_{11}/M_{N_1}$, $\tilde
m_\Delta$, $\tilde m_\Delta^\ell$, $M_\Delta$ and $r$ \footnote{In
  contrast to the pure triplet leptogenesis case
  \cite{Hambye:2005tk} where the determination of the $L$ asymmetry
  depends only on three parameters: $\tilde m_\Delta$, $\tilde
  m_\Delta^\ell$, $M_\Delta$.}.
\section{Results}
\label{sec:res}
The details of the generation of the lepton asymmetry are determined
by the relative size of the different parameters intervening in
eqs. \eqref{eq:relevant-BEQs}, which fix the size of the corresponding
Yukawa, Higgs and gauge reaction densities and also fix the direction
in the ``lepton-Higgs space'' through which the triplet, Higgs and
lepton asymmetries are projected (see
eqs. \eqref{eq:projectors}). Three possible scenarios can be defined
\cite{AristizabalSierra:2011ab}:
\begin{enumerate}[A.]
\item \underline{Purely triplet scalar leptogenesis models:}\\
  The relevant parameters follow the hierarchy $\tilde m_1\ll
  \tilde m_\Delta^\ell, \tilde m_\Delta$. The $L$ asymmetry is
  generated through the processes $\pmb{\Delta}\to \bar\ell\bar\ell$
  or $\pmb{\Delta}\to H H$ and the details strongly depend on whether
  $\tilde m_\Delta^\ell\gg \tilde m_\Delta$, $\tilde m_\Delta^\ell\ll
  \tilde m_\Delta$ or $\tilde m_\Delta^\ell\sim \tilde m_\Delta$.
  Interestingly, when $\tilde m_\Delta^\ell\gg \tilde m_\Delta$ the
  Higgs asymmetry---being weakly washed out---turns out to be large
  and implies a large lepton asymmetry.
\item \underline{Singlet dominated leptogenesis models:}\\
  These scenarios are defined according to $\tilde m_1\gg \tilde
  m_\Delta^\ell, \tilde m_\Delta$ thus leptogenesis is mainly
  determined by $N_1$ dynamics. The relative difference between the
  parameters $\tilde m_\Delta^\ell$ and $\tilde m_\Delta$ determines
  whether either the Higgs asymmetry or the $L$ asymmetry are strongly
  or weakly washed out, thus three cases can be distinguished: $\tilde
  m_\Delta^\ell\gg \tilde m_\Delta$, $\tilde m_\Delta^\ell\ll \tilde
  m_\Delta$ or $\tilde m_\Delta^\ell\sim \tilde m_\Delta$. Each of
  them exhibit different features.
\item \underline{Mixed leptogenesis models:}\\
  In these models the parameters controlling the gauge reaction
  densities strengths are all of the same order i.e.  $\tilde
  m_1\sim \tilde m_\Delta^\ell\sim \tilde m_\Delta$.
\end{enumerate}
\begin{figure}
  \centering
  \includegraphics[width=7cm,height=5.5cm]{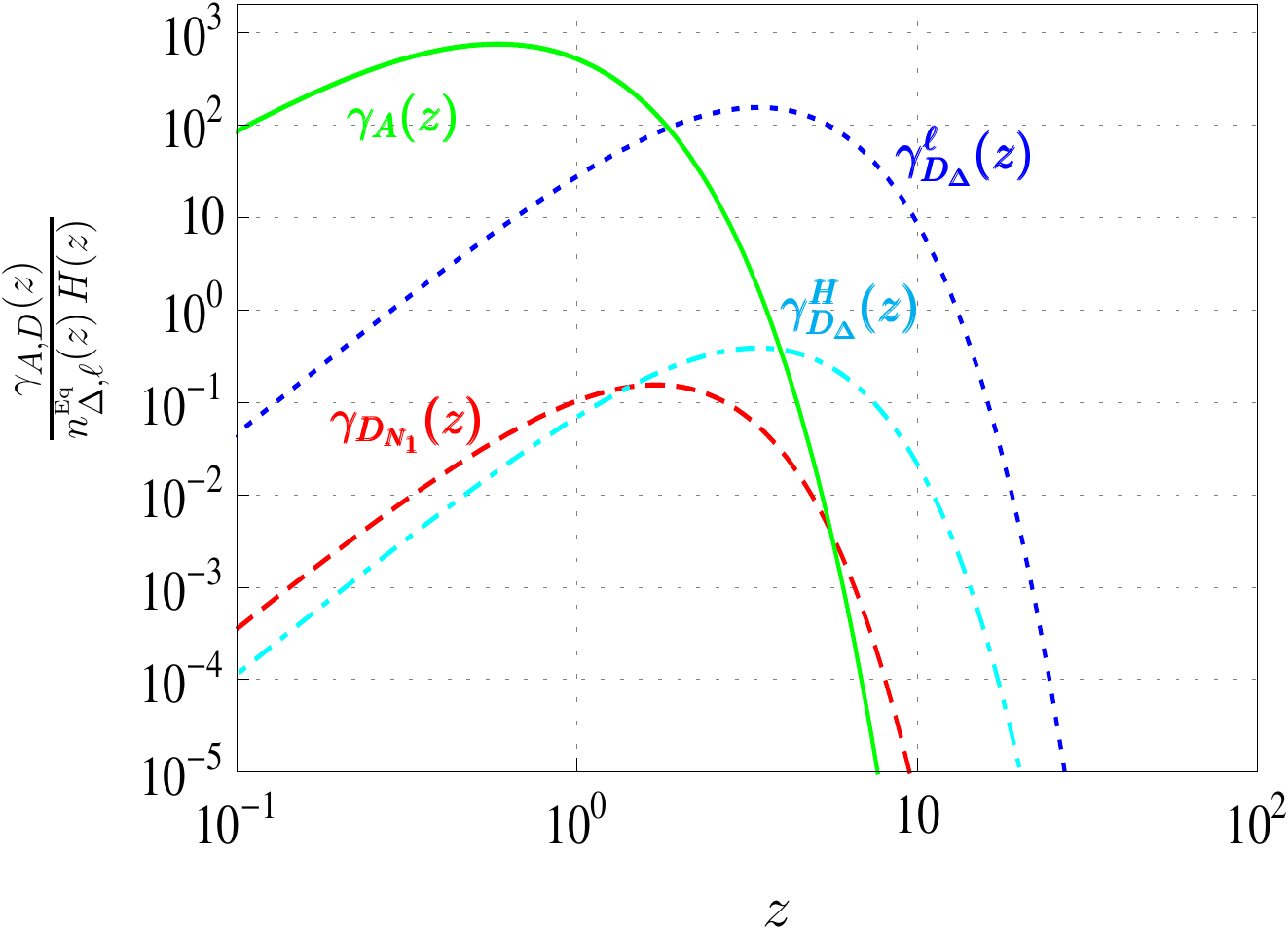}
  \caption{Reaction densities for triplet and RH neutrino processes.}
  \label{fig:reaction-dens}
\end{figure}

\begin{figure}
  \centering
  \includegraphics[width=7cm,height=5.5cm]{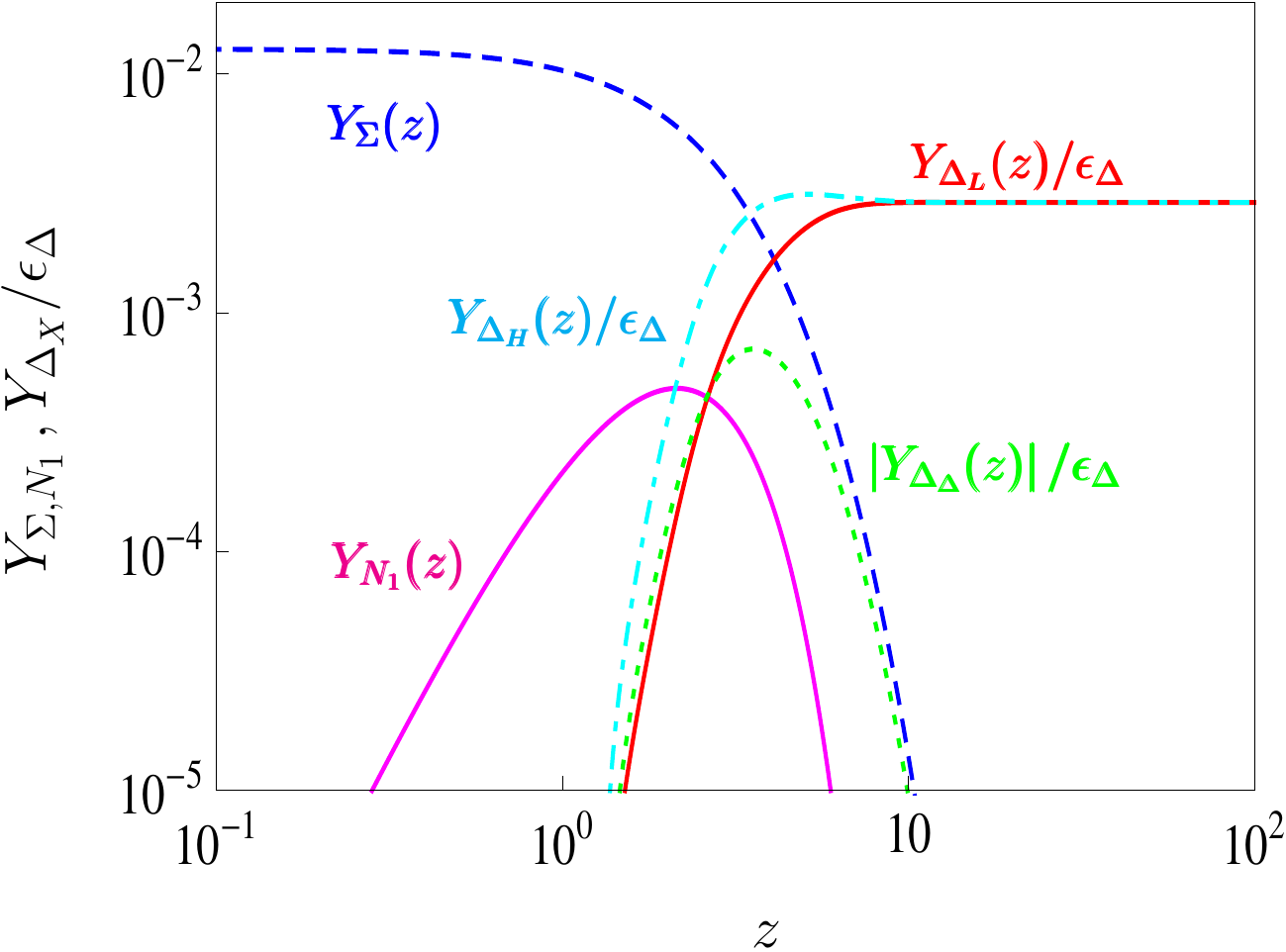}
  \caption{Evolution of the different densities as a function of
    $z=M_\Delta/T$.}
  \label{fig:densities}
\end{figure}

A full discussion of these schemes can be found in
\cite{AristizabalSierra:2011ab}. Here in order to illustrate the main
features of leptogenesis in these scenarios we show in
fig. \ref{fig:reaction-dens} the different gauge reaction densities
obtained by choosing $P_I$=($\tilde m_1$, $\tilde m_\Delta$, $\tilde
m_\Delta^\ell$,$M_\Delta$,$r$) =($10^{-4}$ eV, $10^{-2}$ eV, $10^{-1}$
eV, $10^{10}$ GeV,2) and fixing $\epsilon_{N_1}^\text{tot}=10^{-6}$
and $\epsilon_\Delta=10^{-5}$.  As expected, at high temperature gauge
reactions dominate and start rapidly falling at $z\sim 1$. Above that
value they are overcomed by Higgs related reactions which although
large do not imply a strong washout of the Higgs asymmetry (due to
$\gamma_{D_\Delta}^H\ll \gamma_{D_\Delta}^\ell$) as demonstrated in
fig. \ref{fig:densities}. Indeed the large Higgs asymmetry allows the
development of a large triplet and lepton asymmetry at values slightly
above $z\sim 1$, as required by
condition~\eqref{eq:hypercharge-neutrality}. At higher values the
triplet asymmetry is diluted and transferred to the lepton asymmetry
which accordingly increases, and matches the Higgs asymmetry when the
triplet asymmetry is entirely depleted. This effect (storing a large
asymmetry in certain lepton-Higgs direction) resembles what can happen
in standard leptogenesis when flavor effects are taken into account.
\section{Conclusions}
\label{sec:conc}
We discussed scenarios where leptogenesis takes place due to interplay
between type-I and type-II seesaws. In particular, we have analysed
scenarios where the lepton asymmetry is generated through the CP
violating decays of the lightest RH neutrino and the electroweak
triplet. Working in the one-flavor approximation, it has been shown
that three scenarios for the generation of a baryon asymmetry via
leptogenesis can be identified. Even under this assumption we have
found that in some of the scenarios discussed, the simultaneous
presence of triplet and RH neutrino interactions allows an enhancement
of the lepton asymmetry produced via leptogenesis.




\nocite{*}
\bibliographystyle{elsarticle-num}
\bibliography{stlept-bib}

\begin{thebibliography}{1}
\expandafter\ifx\csname url\endcsname\relax
  \def\url#1{\texttt{#1}}\fi
\expandafter\ifx\csname urlprefix\endcsname\relax\def\urlprefix{URL }\fi
\expandafter\ifx\csname href\endcsname\relax
  \def\href#1#2{#2} \def\path#1{#1}\fi

\bibitem{Hambye:2012fh}
T.~Hambye, {Leptogenesis: beyond the minimal type I seesaw scenario}\href
  {http://arxiv.org/abs/1212.2888} {\path{arXiv:1212.2888}}.

\bibitem{Hambye:2005tk}
T.~Hambye, M.~Raidal, A.~Strumia, {Efficiency and maximal CP-asymmetry of
  scalar triplet leptogenesis}, Phys.Lett. B632 (2006) 667--674.
\newblock \href {http://arxiv.org/abs/hep-ph/0510008}
  {\path{arXiv:hep-ph/0510008}}, \href
  {http://dx.doi.org/10.1016/j.physletb.2005.11.007}
  {\path{doi:10.1016/j.physletb.2005.11.007}}.

\bibitem{Hambye:2003rt}
T.~Hambye, Y.~Lin, A.~Notari, M.~Papucci, A.~Strumia, {Constraints on neutrino
  masses from leptogenesis models}, Nucl.Phys. B695 (2004) 169--191.
\newblock \href {http://arxiv.org/abs/hep-ph/0312203}
  {\path{arXiv:hep-ph/0312203}}, \href
  {http://dx.doi.org/10.1016/j.nuclphysb.2004.06.027}
  {\path{doi:10.1016/j.nuclphysb.2004.06.027}}.

\bibitem{AristizabalSierra:2010mv}
D.~Aristizabal~Sierra, J.~F. Kamenik, M.~Nemevsek, {Implications of Flavor
  Dynamics for Fermion Triplet Leptogenesis}, JHEP 1010 (2010) 036.
\newblock \href {http://arxiv.org/abs/1007.1907} {\path{arXiv:1007.1907}},
  \href {http://dx.doi.org/10.1007/JHEP10(2010)036}
  {\path{doi:10.1007/JHEP10(2010)036}}.

\bibitem{AristizabalSierra:2011ab}
D.~Aristizabal~Sierra, F.~Bazzocchi, I.~de~Medeiros~Varzielas, {Leptogenesis in
  flavor models with type I and II seesaws}, Nucl.Phys. B858 (2012) 196--213.
\newblock \href {http://arxiv.org/abs/1112.1843} {\path{arXiv:1112.1843}},
  \href {http://dx.doi.org/10.1016/j.nuclphysb.2012.01.009}
  {\path{doi:10.1016/j.nuclphysb.2012.01.009}}.

\bibitem{AristizabalSierra:2012js}
D.~Aristizabal~Sierra, I.~de~Medeiros~Varzielas, {The role of lepton flavor
  symmetries in leptogenesis}, Fortschritte der Physik (Progress of
  physics)\href {http://arxiv.org/abs/1205.6134} {\path{arXiv:1205.6134}},
  \href {http://dx.doi.org/10.1002/prop.201200122}
  {\path{doi:10.1002/prop.201200122}}.

\bibitem{Hambye:2003ka}
T.~Hambye, G.~Senjanovic, {Consequences of triplet seesaw for leptogenesis},
  Phys.Lett. B582 (2004) 73--81.
\newblock \href {http://arxiv.org/abs/hep-ph/0307237}
  {\path{arXiv:hep-ph/0307237}}, \href
  {http://dx.doi.org/10.1016/j.physletb.2003.11.061}
  {\path{doi:10.1016/j.physletb.2003.11.061}}.

\bibitem{Nardi:2006fx}
E.~Nardi, Y.~Nir, E.~Roulet, J.~Racker, {The Importance of flavor in
  leptogenesis}, JHEP 0601 (2006) 164.
\newblock \href {http://arxiv.org/abs/hep-ph/0601084}
  {\path{arXiv:hep-ph/0601084}}, \href
  {http://dx.doi.org/10.1088/1126-6708/2006/01/164}
  {\path{doi:10.1088/1126-6708/2006/01/164}}.

\bibitem{Abada:2006fw}
A.~Abada, S.~Davidson, F.-X. Josse-Michaux, M.~Losada, A.~Riotto, {Flavor
  issues in leptogenesis}, JCAP 0604 (2006) 004.
\newblock \href {http://arxiv.org/abs/hep-ph/0601083}
  {\path{arXiv:hep-ph/0601083}}, \href
  {http://dx.doi.org/10.1088/1475-7516/2006/04/004}
  {\path{doi:10.1088/1475-7516/2006/04/004}}.

\end{thebibliography}







\end{document}